\documentclass{aa}
\usepackage{txfonts}
\usepackage{graphicx}
\usepackage{subfigure}
\usepackage{natbib}
\bibpunct{(}{)}{;}{a}{}{,}

\def\chandra{{\it Chandra~}}

\def\swift{{\it Swift~}}

\def\xmm{{XMM-Newton~}}
\def\xmmk{{XMM-Newton}}

\def\m31{{M~31}}
\def\nova{{M31N~2007-06b~}}
\def\novak{{M31N~2007-06b}}
\def\gsss{{1E~1339.8+2837~}}
\def\gsssk{{1E~1339.8+2837}}

\newcommand{\nh}{\hbox{$N_{\rm H}$}~}
\newcommand{\hcm}[1]{$\times 10^{#1}$ cm$^{-2}$}

\newcommand{\ergs}[1]{$\times 10^{#1}$ \hbox{erg s$^{-1}$}}
\newcommand{\oergs}[1]{$10^{#1}$ erg s$^{-1}$}
\newcommand{\cts}[1]{$\times 10^{#1}$ ct s$^{-1}$}

\newcommand{\tpower}[1]{$\times 10^{#1}$}
\newcommand{\power}[1]{$10^{#1}$}

\begin{document}

\title{The first two transient supersoft X-ray sources in \m31 globular clusters and the connection to classical novae\thanks{Partly 
   based on observations with \xmmk, an ESA Science Mission with instruments and contributions directly funded by ESA Member States and NASA}}

\author{M.~Henze\inst{1}
	\and W.~Pietsch\inst{1}
	\and F.~Haberl\inst{1}
	\and G.~Sala\inst{1,2}
	\and R.~Quimby\inst{3}\thanks{current address: California Institute of Technology, Pasadena, CA 91125, USA}
	\and M.~Hernanz\inst{4}
	\and M.~Della Valle\inst{5,6,7}
	\and P.~Milne\inst{8}
	\and G.G.~Williams\inst{8}
	\and V.~Burwitz\inst{1}
	\and J.~Greiner\inst{1}
	\and H.~Stiele\inst{1}
	\and D.H.~Hartmann\inst{9}
	\and A.K.H.~Kong\inst{10}
	\and K.~Hornoch\inst{11}
}

\institute{Max-Planck-Institut f\"ur extraterrestrische Physik, 
	D-85748 Garching, Germany\\
	email: mhenze@mpe.mpg.de
	\and Departament de F\'isica i Enginyeria Nuclear, EUETIB (UPC/IEEC), Comte d'Urgell 187, 08036 Barcelona, Spain
	\and University of Texas, Austin TX, 78712, USA
	\and Institut de Ci\`encies de l'Espai (CSIC-IEEC), Campus UAB, Fac. Ci\`encies, E-08193 Bellaterra, Spain
	\and European Southern Observatory (ESO), D-85748 Garching, Germany
	\and INAF-Napoli, Osservatorio Astronomico di Capodimonte, Salita Moiariello 16, I-80131 Napoli, Italy
	\and International Centre for Relativistic Astrophysics, Piazzale della Repubblica 2, I-65122 Pescara, Italy
	\and Steward Observatory, 933 North Cherry Avenue, Tucson, AZ 85721, USA
	\and Department of Physics and Astronomy, Clemson University, Clemson, SC 29634-0978, USA
	\and Institute of Astronomy and Department of Physics, National Tsing Hua University, Hsinchu, Taiwan
	\and Astronomical Institute, Academy of Sciences, CZ-251 65 Ond\v{r}ejov, Czech Republic
}

\date{Received 10 October 2008 / Accepted 21 April 2009}

\abstract
{Classical novae (CNe) have been found to represent the major class of supersoft X-ray sources (SSSs) in our neighbour galaxy \m31.}
{We determine the properties and evolution of the two first SSSs ever discovered in the \m31 globular cluster (GC) system.}
{We have used \xmmk, \chandra and \swift observations of the centre region of \m31 to discover both SSSs and to determine their X-ray light curves and spectra. We performed detailed analysis of \xmm EPIC PN spectra of the source in Bol 111 (SS1) using blackbody and NLTE white dwarf (WD) atmosphere models. For the SSS in Bol 194 (SS2) we used optical monitoring data to search for an optical counterpart.}
{Both GC X-ray sources were classified as SSS. We identify SS1 with the CN \nova recently discovered in the \m31 GC Bol 111. For SS2 we did not find evidence for a recent nova outburst and can only provide useful constraints on the time of the outburst of a hypothetical nova.}
{The only known CN in a \m31 GC can be identified with the first SSS found in a \m31 GC. We discuss the impact of our observations on the nova rate for the \m31 GC system.}


\titlerunning{Supersoft X-ray sources in M\,31 globular clusters}

\maketitle

%
%
\section{Introduction}
%

Supersoft X-ray sources (SSSs) are a class of X-ray sources that were first characterised based on ROSAT observations \citep[see e.g.][]{1991A&A...246L..17G}. These sources show extremely soft X-ray spectra, with little or no radiation at energies above 1 keV \citep[see e.g.][]{1998A&A...332..199P}, that can be described by blackbody temperatures typically in the range of 15-80 eV \citep[see][and references therein]{1997ARA&A..35...69K}. SSSs were originally believed to be hydrogen burning white dwarfs (WDs) in binary systems, where the WD steadily burns hydrogen rich matter accreted from its companion star \citep{1997ARA&A..35...69K}. However, the class of SSSs is quite inhomogeneous. Prototypical sources are, on the one hand, CAL 83 \citep[see][and references therein]{1998A&A...332..199P} and CAL 87 \citep[see][and references therein]{1997A&A...323L..33P}, both of which are located in the Large Magellanic Cloud. These objects show eclipses (CAL 87) or rare X-ray off states (CAL 83) \citep{1997ARA&A..35...69K}, but are rather permanent SSSs. On the other hand, \citet{2005A&A...442..879P} found out that classical novae (CNe) represent the major class of SSSs in our neighbour galaxy \m31 \citep[distance 780 kpc,][]{1998AJ....115.1916H,1998ApJ...503L.131S}. CNe can appear as luminous ($L\sim$\oergs{38}) transient SSSs that seem to go through a single outburst that can last from months to several years \citep{2007A&A...465..375P}.

Classical novae (CNe) are thermonuclear explosions on the surface of WDs in cataclysmic binaries that result from the transfer of matter from the companion star to the WD. The transferred hydrogen-rich matter accumulates on the surface of the WD until hydrogen ignition starts a thermonuclear runaway in the degenerate matter of the WD envelope. The resulting expansion of the hot envelope can cause the brightness of the WD to rise by more than nine magnitudes within a few days, and mass to be ejected at high velocities \citep[see][and references therein]{2005ASPC..330..265H,1995cvs..book.....W}. However, a fraction of the hot envelope can remain in steady hydrogen burning on the surface of the WD \citep{1974ApJS...28..247S,2005A&A...439.1061S}, powering a supersoft X-ray source that can be observed directly once the ejected envelope becomes sufficiently transparent \citep{1989clno.conf...39S,2002AIPC..637..345K}.

The duration of the supersoft phase is related to the amount of H-rich matter that is {\it not} ejected and also depends on the luminosity of the white dwarf. More massive WDs need to accrete less matter to initiate the thermonuclear runaway, because of their higher surface gravity \citep{1998ApJ...494..680J}. In general, more massive WDs retain less accreted mass after the explosion, although this also depends on the accretion rate, and reach larger luminosities \citep{2005ApJ...623..398Y}. Thus, the duration of the SSS state is inversely related to the mass of the WD \citep{2005A&A...439.1061S,1998ApJ...503..381T}. In turn, the transparency requirement mentioned above implies that the time of appearance of the SSS is determined by the fraction of mass ejected in the outburst \citep{2006ApJS..167...59H}. X-ray lightcurves therefore provide important clues on the physics of the nova outburst, addressing the key question whether a WD accumulates matter over time to become a potential
progenitor for a type Ia supernova (SN-Ia).

Due to its proximity to the Galaxy and its moderate Galactic foreground absorption \citep[\nh = 0.7\hcm{21},][]{1992ApJS...79...77S}, \m31 is a unique target for CN surveys. Starting with \citet{1929ApJ....69..103H} the majority of these surveys that were conducted in the past \citep[see e.g.][and references therein]{2008A&A...477...67H} classified optical transients as novae just by their lightcurve. Following the steep rise, the luminosity of different CNe declines with different speed, which allows a phenomenological classification of these objects by their speed class \citep{1964gano.book.....P}. Eventually, novae in \m31 will fade back to invisibility, since the binary systems in quiescence are too faint to be observed at the distance of \m31. Only recently, nova monitoring programs for \m31 were established, that include fast data analysis and that therefore provide the possibility to conduct follow-up spectroscopy \citep[see e.g.][]{2008ATel.1654....1H,2008ATel.1703....1M} and to classify CNe within the system of \citet{1992AJ....104..725W}.

Almost all optical surveys for CNe in \m31 that were conducted in the past searched for suddenly appearing objects that have not been visible before and fade back to invisibility in days to weeks. This condition is certainly not fulfilled by CNe in relatively bright GCs, where the optical background light of the GC itself makes a photometric discovery of a nova outburst much more complicated. Therefore, the connection of CNe to SSSs in X-rays provides a useful possibility to detect CNe in GCs. Novae in GCs are rare. There were just two sources known so far that likely fit this definition. One of them was seen in the Galactic GC M\,80 whereas the second nova was found in a GC of the galaxy M\,87 \citep[see][and references therein]{2004ApJ...605L.117S}. According to \citet{2004ApJ...605L.117S} a third candidate (nova 1938 in the galactic GC M\,14) is less likely to be a GC nova.  Recently, \citet{2007ApJ...671L.121S} reported the very first nova found in a \m31 GC (\novak).

Independent from their connection to CNe, SSSs in GCs are rare objects. There was just one SSS known in GC up to now: the transient \gsss in the Galactic GC M\,3 (NGC 5272) \citep{1999PASJ...51..519D,1995A&A...300..732V}. \gsss was first discovered with the Einstein satellite with a hard X-ray spectrum and a low luminosity of 4 \ergs{33} \citep{1983ApJ...267L..83H} but showed a very soft high luminosity state during 1991-92 when observed with ROSAT \citep{1999PASJ...51..519D}. \citet{1999PASJ...51..519D} give for \gsss a blackbody temperature of $kT \sim$ 36 eV and a luminosity of $\sim$ \oergs{35}, which is significantly lower than the observed peak luminosities of other SSS, including the two sources discussed in this work. They discuss this object as a cataclysmic variable (CV) system that may be a dwarf nova including a massive WD. The CV interpretation is supported by \citet{2004ApJ...611..413E} who used Hubble Space Telescope observations to discover an optical counterpart to \gsssk. They suggest that magnetically channeled accretion could explain the peculiarities of \gsssk.

Recently, we reported, based on preliminary data analysis \citep{2007ATel.1294....1P}, the very first SSS (hereafter SS1) that was found in a \m31 GC. During the same observing run where we found this source we detected a second luminous and previously unknown SSS (hereafter SS2) in another \m31 GC \citep{2007ATel.1296....1H,2007ATel.1306....1H}. Both sources are transients and will be examined in detail in this work.

The structure of the paper is as follows: In Sect.\,2 we describe our X-ray observations and the properties of the two SSSs. The optical data available for Bol 194 and the data analysis are presented and interpreted in Sect.\,3. Finally, in Sect.\,4 we summarise our results and discuss the impact that two GC novae within one year would have on the nova rate for the \m31 GC system.

%
%
\section{X-ray observations and data analysis}
\label{sec:obs_xray}
\begin{figure}
	\resizebox{\hsize}{!}{\includegraphics[angle=270]{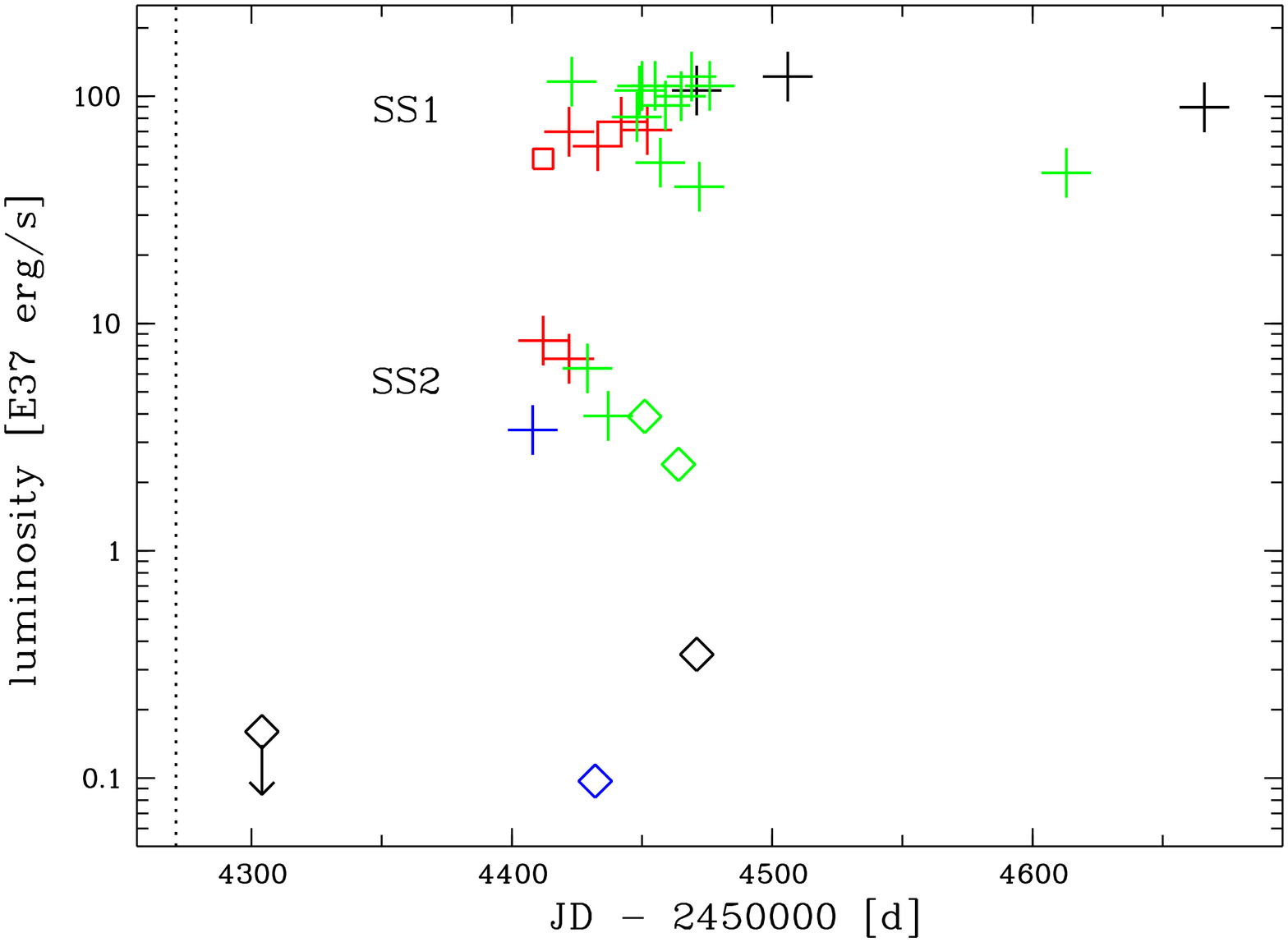}}
	\caption{X-ray light curves of SS1 and SS2 (0.2 - 1.0 keV) obtained from \xmm (\textbf{black}), \chandra HRC-I (\textbf{red}), \chandra ACIS-S (\textbf{blue}), and \swift XRT (\textbf{green}) (see also Tables\,\ref{table:ss1_xray},\ref{table:ss2_xray}). The upper limit for SS1 is indicated by a \textbf{down-pointing arrow}, the upper limits for SS2 are marked by \textbf{open lozenges}. An \textbf{open square} indicates the lower limit luminosity for SS1 on \chandra ObsID 8526. \textbf{Crosses} symbolise luminosities of detections. Error bars are not shown, since the (statistical) errors are mostly smaller than the size of the symbols. There are no errors given by \citet{2007ATel.1328....1G} for the ACIS-S observations. The vertical dotted line indicates the day of the first detection of nova \nova in the optical \citep{2007ApJ...671L.121S}.}
	\label{fig:xray_light}
\end{figure}

\begin{figure}
	\resizebox{\hsize}{!}{\includegraphics[angle=0]{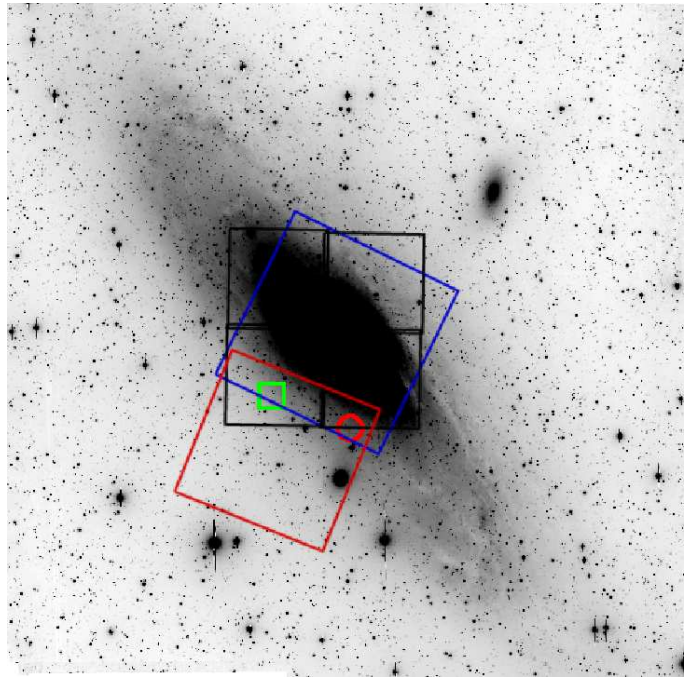}}
	\caption{ROTSE-III \m31 central field (see Sec.\,\ref{sec:obs_optical}). Overlaid are the four Super-LOTIS fields (black squares), the \chandra HRC-I field for ObsID 8527 (big blue square) and the \xmm field for ObsID 0511380201 (big red square). Indicated are the position of Bol 111 (red circle) and Bol 194 (green square).}
	\label{fig:rotse_m31}
\end{figure}

%
In the context of the \xmmk/\chandra \m31 nova monitoring project\footnote{http://www.mpe.mpg.de/$\sim$m31novae/xray/index.php} we obtained five 20 ks \chandra HRC-I observations of M\,31 starting in November 2007. We detected two new sources in the M\,31 GCs Bol 111 and Bol 194 \citep{2004A&A...416..917G}. For both sources follow-up ToO observations with the \swift X-ray Telescope (XRT) were requested \citep[e.g.][]{2007ATel.1294....1P,2007ATel.1333....1K} as well as additional \chandra ACIS-S observations by another group \citep{2007ATel.1328....1G}. We also used two observations obtained within the \xmm \m31 large survey project\footnote{\xmm Large Program: The X-ray source population of the Andromeda galaxy \m31; PI: W. Pietsch\\ see http://www.mpe.mpg.de/xray/research/normal\_galaxies/m31/lp.php}. Details on the observations used are given in Tables\,\ref{table:ss1_xray} and \ref{table:ss2_xray} for the sources in Bol 111 and Bol 194, respectively. The tables list the telescopes and instruments used, the observation identifications (ObsIDs), the dates as well as source count rates and luminosities or upper limits, respectively. X-ray light curves of both sources are shown in Fig.\,\ref{fig:xray_light}.
Figure\,\ref{fig:rotse_m31} shows the fields of view of \chandra and \xmm observations with representative positions.

We analysed all available observations using mission dependent source detection software as well as the HEAsoft package v6.3, including the spectral analysis software XSPEC v12.3.1. In all our XSPEC models we used the T\"ubingen-Boulder ISM absorption (\texttt{TBabs} in XSPEC) model together with the photoelectric absorption cross-sections from \citet{1992ApJ...400..699B} and ISM abundances from \citet{2000ApJ...542..914W}. For the individual telescopes we applied the following data reduction techniques:

The \swift XRT data were analysed using the HEAsoft XIMAGE package (version 4.4) with the \texttt{sosta} command (source statistics) for estimations of count rates. We took into account the XRT PSF of the sources that we computed with the command \texttt{psf}, as well as exposure maps that were created with the XRT software task \texttt{xrtexpomap} within XIMAGE. Astrometry was done using the HEAsoft routine \texttt{xrtcentroid}.

For the \xmm data we applied a background screening and used the XMMSAS v6.6 tasks \texttt{eboxdetect} and \texttt{emldetect} to detect sources in the image and perform astrometry and photometry. For computing upper limits we added an artificial detection at the position of the source to the \texttt{eboxdetect} list. This list was used as input for an \texttt{emldetect} run (with fixed positions and likelihood threshold of zero) that derived the observed flux and upper limit for all objects in the list. We give 3 times the background flux as the 3$\sigma$ upper limit. To obtain astrometrically-corrected positions we selected optical sources, with proper-motion corrected positions, from the USNO-B1.0 catalogue \citep{2003AJ....125..984M}, and checked that only one optical source was in the error circle of the corresponding X-ray source. We only accepted sources correlating with globular clusters from the Revised Bologna Catalogue \citep[V.3.4, January 2008;][]{2004A&A...416..917G,2005A&A...436..535G,2006A&A...456..985G,2007A&A...471..127G} or with foreground stars, characterized by their optical to X-ray flux ratio \citep{1988ApJ...326..680M} and their hardness ratio \citep[see][]{2008A&A...480..599S}. We then used the XMMSAS task {\tt eposcorr} to derive the offset of the X-ray aspect solution.

We reduced the \chandra observations with the CIAO v3.4 (Chandra Interactive Analysis of Observations) software package. The source detection was done with the CIAO tool \texttt{wavedetect}. Similar to the procedure described by \citet{2009arXiv0903.2062E} for the \chandra COSMOS survey, an adapted version of the XMMSAS tool \texttt{emldetect} was used to estimate background and exposure corrected fluxes and count rates for the detected sources. Both sources are located near the edge of the HRC-I field of view and therefore both, photometry and astrometry, suffer from relatively large errors.

%
\begin{table*}[ht]
\caption{X-ray observations of SS1 in Bol 111.}
\label{table:ss1_xray}
\begin{center}
\begin{tabular}{lrrrrrr}\hline\hline \noalign{\smallskip}
	Telescope/Instrument$^a$ & ObsID & Exp. time$^b$ & Date$^c$ & Offset$^d$ & Count Rate$^e$ & L$_{0.2-1.0}$ $^e$\\
	 & & [ks] & [UT] & [d] & [ct s$^{-1}$] & [erg s$^{-1}$]\\ \hline \noalign{\smallskip}
	\xmm EPIC PN & 0505760201 & 49.2 & 2007-07-22.55 & 33 & $<$ 1.1 \tpower{-3} &  $<$1.4 \tpower{36} \\ \hline \noalign{\smallskip}
	\chandra HRC-I & 8526 & 18.7 & 2007-11-07.64 & 141 & $>$ (1.9 $\pm$ 0.4) \tpower{-2} & $>$(5.3 $\pm$ 1.1) \tpower{38}\\
	\chandra HRC-I & 8527 & 20.0 & 2007-11-17.76 & 151 & (2.5 $\pm$ 0.2) \tpower{-2} &  (7.0 $\pm$ 0.7) \tpower{38}\\
	\chandra HRC-I & 8528 & 20.0 & 2007-11-28.79 & 162 & (2.2 $\pm$ 0.2) \tpower{-2} &  (6.0 $\pm$ 0.6) \tpower{38} \\
	\chandra HRC-I & 8529 & 18.9 & 2007-12-07.57 & 171 & (2.8 $\pm$ 0.3) \tpower{-2} &  (7.7 $\pm$ 0.7) \tpower{38}\\
	\chandra HRC-I & 8530 & 19.9 & 2007-12-17.49 & 181 & (2.5 $\pm$ 0.3) \tpower{-2} &  (7.1 $\pm$ 0.7) \tpower{38}\\ \hline \noalign{\smallskip}
	\swift XRT & 00031017001/2 & 7.1 & 2007-11-18.40 & 152 & (1.18 $\pm$ 0.15) \tpower{-2} &  (11.6 $\pm$ 1.4) \tpower{38}\\
	\swift XRT & 00031017003 & 3.0 & 2007-11-13.02 &  177 & (0.8 $\pm$ 0.2) \tpower{-2} &  (8.1 $\pm$ 1.9) \tpower{38}\\
	\swift XRT & 00031017004 & 3.0 & 2007-12-14.02 &  178 & (1.1 $\pm$ 0.2) \tpower{-2} &  (10.6 $\pm$ 2.2) \tpower{38}\\
	\swift XRT & 00031017005 & 3.2 & 2007-12-15.03 &  179 & (1.1 $\pm$ 0.2) \tpower{-2} &  (11.1 $\pm$ 2.1) \tpower{38}\\
	\swift XRT & 00031017006 & 2.2 & 2007-12-20.25 &  184 & (1.1 $\pm$ 0.3) \tpower{-2} &  (11.1 $\pm$ 2.6) \tpower{38}\\
	\swift XRT & 00031017007 & 2.1 & 2007-12-22.39 &  186 & (0.5 $\pm$ 0.2) \tpower{-2} &  (5.1 $\pm$ 1.9) \tpower{38}\\
	\swift XRT & 00031017008 & 2.3 & 2007-12-24.33 &  188 & (0.9 $\pm$ 0.2) \tpower{-2} &  (9.1 $\pm$ 2.4) \tpower{38}\\
	\swift XRT & 00031017009 & 2.3 & 2007-12-30.15 &  194 & (1.0 $\pm$ 0.2) \tpower{-2} &  (10.0 $\pm$ 2.4) \tpower{38}\\
	\swift XRT & 00031017010 & 2.0 & 2008-01-03.44 &  198 & (1.2 $\pm$ 0.3) \tpower{-2} &  (12.2 $\pm$ 2.8) \tpower{38}\\
	\swift XRT & 00031017011 & 1.9 & 2008-01-06.25 &  201 & (0.4 $\pm$ 0.3) \tpower{-2} &  (4.0 $\pm$ 2.8) \tpower{38}\\
	\swift XRT & 00031017012 & 1.7 &2008-01-10.00 &  205 & (1.1 $\pm$ 0.3) \tpower{-2} &  (11.1 $\pm$ 1.3) \tpower{38}\\ \hline \noalign{\smallskip}
	\xmm EPIC PN & 0511380201 & 23.0 & 2008-01-05.99 & 200 & (8.2 $\pm$ 0.2) \tpower{-2} & (10.6 $\pm$ 0.2) \tpower{38} \\
	\xmm EPIC PN & 0511380601 & 24.0 & 2008-02-09.31 & 235 & (7.5 $\pm$ 0.2) \tpower{-2} & (12.2 $\pm$ 0.4) \tpower{38}\\
	\swift XRT & 00037718001 & 4.8 & 2008-05-26.29 &  342 & (0.5 $\pm$ 0.1) \tpower{-2} &  (4.6 $\pm$ 1.3) \tpower{38}\\ \hline \noalign{\smallskip}
	\xmm EPIC PN & 0560180101 & 17.4 & 2008-07-18.26 & 395 & (3.0 $\pm$ 0.2) \tpower{-2} & (8.9 $\pm$ 0.5) \tpower{38}\\ \hline
\end{tabular}
\end{center}
\noindent
Notes:\hspace{0.3cm} $^a $: Telescope and instrument used for observation.\\
\hspace*{1.1cm} $^b $: Dead time corrected exposure time of the observation.\\
\hspace*{1.1cm} $^c $: Start date of the observation.\\
\hspace*{1.1cm} $^d $: Time in days after the discovery of nova \nova in the optical \citep{2007ApJ...671L.121S} on 2007 June 19.38.\\
\hspace*{1.5cm} (JD = 2454271).\\
\hspace*{1.1cm} $^e $: Source count rates, X-ray luminosities (unabsorbed, blackbody fit, 0.2 - 1.0 keV) and upper limits were estimated according\\
\hspace*{1.5cm}to Sect.\,\ref{sect:sss_bol111}. For \chandra ObsID 8526 the source is right on the detector edge, therefore we give lower luminosity limits.\\
\end{table*}
%

%
\begin{table*}[ht]
\caption{X-ray observations of SS2 in Bol 194}
\label{table:ss2_xray}
\begin{center}
\begin{tabular}{lrrrrrr}\hline\hline \noalign{\smallskip}
	Telescope/Instrument$^a$ & ObsID & Exp. time$^b$ & Date$^c$ & Offset$^d$ & Count Rate$^e$ & Luminosity$^e$\\ 
	 & & [ks] & [UT] & [d] & [ct s$^{-1}]$ & [erg s$^{-1}$]\\ \hline \noalign{\smallskip}
	\xmm EPIC PN & 0505760201 & 49.2 & 2007-07-22.55 & 0 & $<$ 2.9 \tpower{-3} &  $<$ 1.6 \tpower{36}\\
	\chandra ACIS-S &  8186 & 5.0 & 2007-11-03.18 & 104 & 2.1 \tpower{-2} &  3.4 \tpower{37}\\
	\chandra HRC-I & 8526 & 18.7 & 2007-11-07.64 & 108 & (3.55 $\pm$ 0.25) \tpower{-2} &  (8.41 $\pm$ 0.59) \tpower{37}\\
	\chandra HRC-I & 8527 & 20.0 & 2007-11-18.76 & 118 & (2.95 $\pm$ 0.25) \tpower{-2} &  (7.00 $\pm$ 0.60) \tpower{37}\\
	\swift XRT & 00031027001 & 7.3 & 2007-11-24.34 & 125 & (8.0 $\pm$ 1.2) \tpower{-3} & (6.35 $\pm$ 1.04) \tpower{37}\\
	\chandra ACIS-S & 8187 & 5.0 & 2007-11-27.16 & 128 & $<$ 6 \tpower{-4} &  $<$ 9.7 \tpower{35} \\
	\swift XRT & 00031027002/3 & 4.7 & 2007-12-02.64 & 133 & (3.2 $\pm$ 1.0) \tpower{-3} & (3.92 $\pm$ 1.22) \tpower{37}\\
	\swift XRT & 00031027004 & 3.9 & 2007-12-16.77 & 147 & $<$ 3.2 \tpower{-3} &   $<$ 3.9 \tpower{37}\\
	\swift XRT & 00031027005 & 4.0 & 2007-12-30.02 & 160 & $<$ 2.0 \tpower{-3} &   $<$ 2.4 \tpower{37}\\
	\xmm EPIC PN & 0511380201 & 22.8 & 2008-01-05.99 & 167 & $<$ 5.1 \tpower{-3} & $<$ 3.5 \tpower{36}\\ 
\hline
\end{tabular}
\end{center}
\noindent
Notes:\hspace{0.3cm} $^a$: Telescope and instrument used for observation. ACIS-S count rates are from \citet{2007ATel.1328....1G}.\\
\hspace*{1.1cm} $^b $: Dead time corrected exposure time of the observation.\\
\hspace*{1.1cm} $^c$: Start date of the observation.\\
\hspace*{1.1cm} $^d $: Time in days after the last X-ray non-detection of SS2 on 2007 July 22.55 (JD = 2454304).\\
\hspace*{1.1cm} $^e $: Source count rates, X-ray luminosities (unabsorbed, blackbody fit, 0.2 - 1.0 keV) and upper limits were estimated\\
\hspace*{1.5cm}according to Sect.\,\ref{sect:sss_bol194}.\\
\end{table*}
%

%

%

\begin{figure}
	\subfigure[Bol 111]{\includegraphics[scale=.23, angle=0]{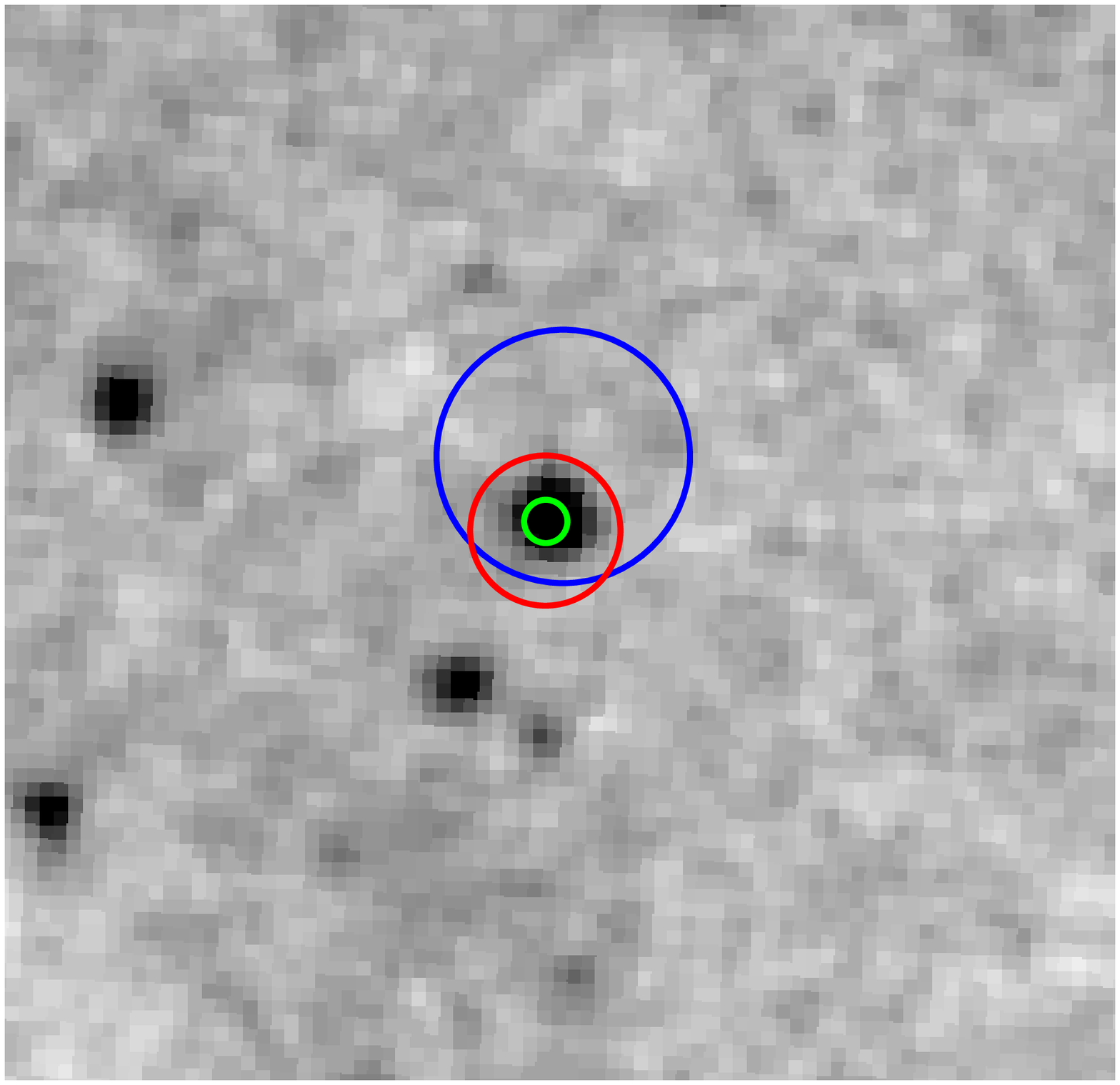}}\,
	\subfigure[Bol 194]{\includegraphics[scale=.23, angle=0]{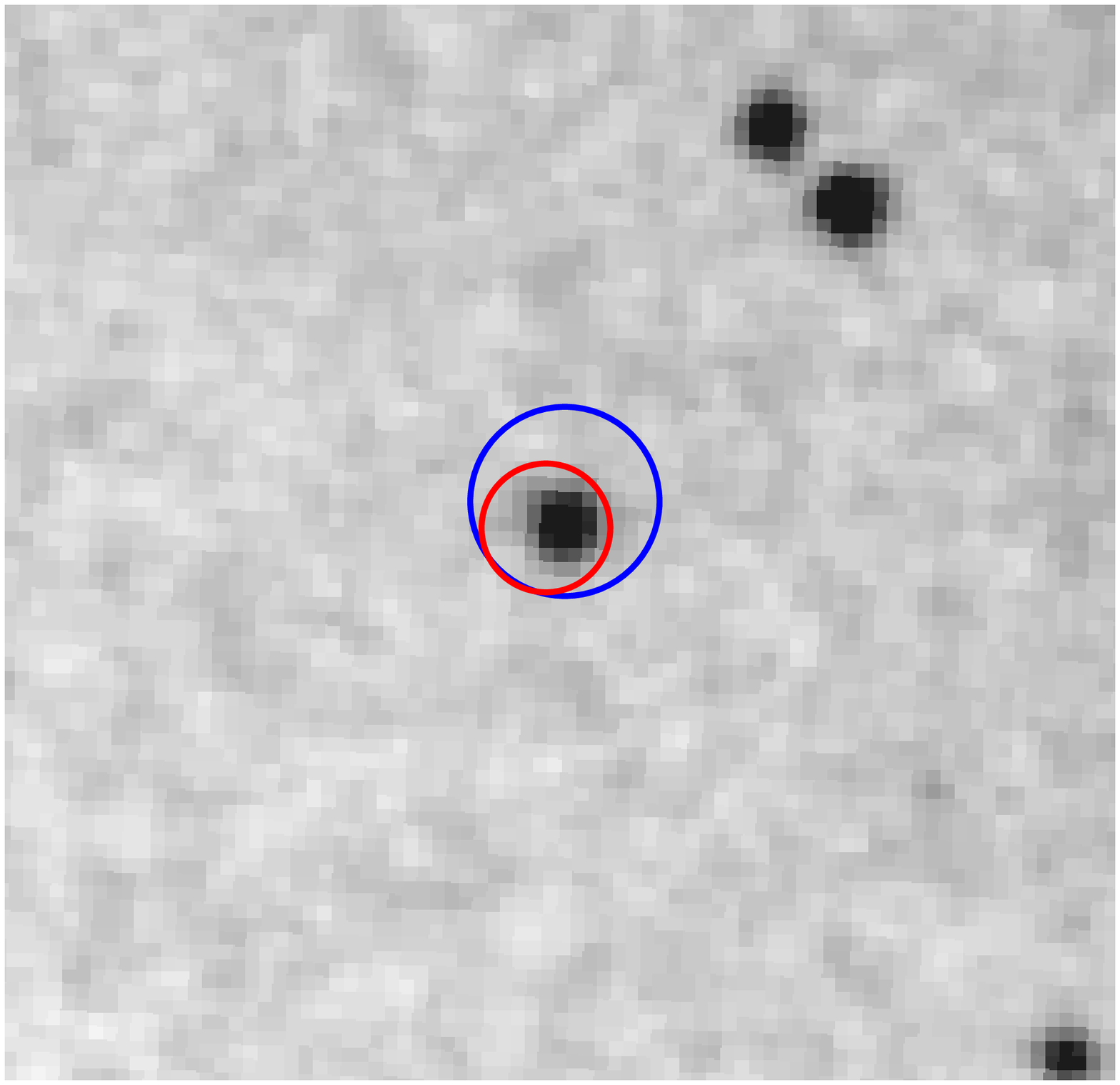}}
	\caption{X-ray position error circles for SS1 and SS2 as described in Sect.\,\ref{sect:sss_bol111} and Sect.\,\ref{sect:sss_bol194} with respect to Bol 111 and Bol 194. \textbf{Blue/big}: \chandra HRC-I, \textbf{red/medium}: \swift XRT, \textbf{green/small}: \xmm PN. Underlying optical image: $5\arcmin \times 5\arcmin$ DSS POSS-II Red.}
	\label{fig:xray_pos_err}
\end{figure}

\subsection{Supersoft source in Bol 111}
\label{sect:sss_bol111}
In our first \chandra observation of the AO6 monitoring campaign, starting on 2007-11-07.64 UT (ObsID 8526), we detected a new source (SS1) at the very edge of the HRC-I field of view \citep{2007ATel.1294....1P}. SS1 remained active during the following \chandra monitoring observations (ObsIDs 8527-30) in 2007 November and December. We followed the light curve of SS1 with \swift ToO observations (ObsIDs 00031017001-12, 2007 November - 2008 January) and also found the source to be still visible in our \xmm \m31 monitoring observations (ObsIDs 0511380201 and 0511380601, 2008 January - February). In Table\,\ref{table:ss1_xray} we present details on all observations. Due to the location of SS1 near the edge of the \chandra HRC-I field of view during all observations (see Fig.\,\ref{fig:rotse_m31} for the location of the \chandra field and the source), we used the \xmm observations to perform precise astrometry. The offset-corrected \xmm position of the source was determined to be RA(J2000) 00:42:33.21, Dec(J2000) +41:00:26.1 with a $3 \sigma$ error of $1\,\farcs6$, including the uncertainty of the offset correction. These coordinates are in good agreement (distance = $0\,\farcs5$), within the errors, with the position of the M\,31 GC Bol 111 \citep[][00:42:33.16, +41:00:26.1]{2004A&A...416..917G}. Therefore, we assume that SS1 is situated within the globular cluster. Note, that no X-ray source was previously known in this GC. See Fig.\,\ref{fig:xray_pos_err} for a visualisation of the agreement of optical (DSS POSS-II Red) and X-ray positions.

To perform spectral analysis of SS1 we used \xmm observations obtained on 2008-01-05.99 UT and 2008-02-09.31 UT (ObsIDs 0511380201 and 0511380601). We extracted the spectra of SS1 from both observations and fitted them simultaneously in order to increase statistics.  We used data from \xmmk's PN detector because of the better sensitivity of PN in the soft band compared to both MOS detectors. The spectra were extracted using XMMSAS task \texttt{evselect}. For both spectra only single-pixel events (PATTERN = 0) were selected.

\begin{figure}
	\resizebox{\hsize}{!}{\includegraphics[angle=270,clip]{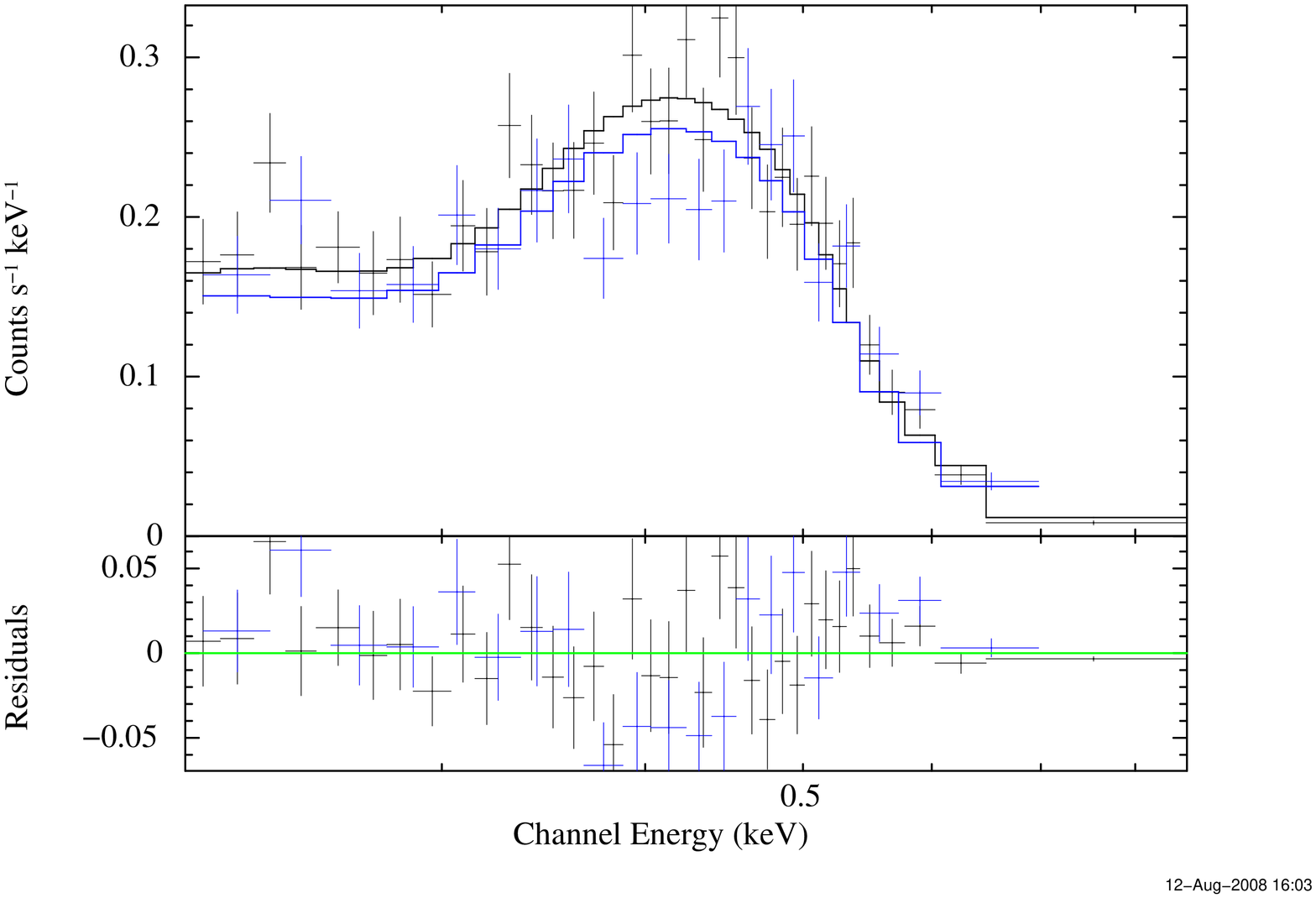}}
	\caption{\xmm EPIC PN spectra of SS1 (\textbf{crosses}) from observations 0511380201 (\textbf{black}) and 0511380601 (\textbf{blue}) fitted with an absorbed blackbody (\textbf{solid lines}).}
	\label{fig:spec_bb_bol111}
\end{figure}

\begin{figure}
	\resizebox{\hsize}{!}{\includegraphics[angle=90]{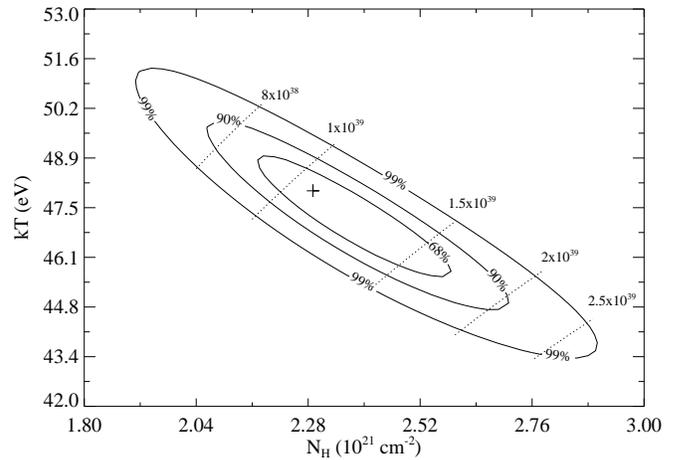}}
	\caption{Column density (\nh) - temperature (kT) contours inferred from the blackbody fit to the \xmm EPIC PN spectra of SS1 (see Fig.\,\ref{fig:spec_bb_bol111}). Normalisation has been adjusted. Indicated are the formal best fit parameters (\textbf{cross}) and the lines of constant X-ray luminosity (0.2-1.0 keV, \textbf{dotted lines}).}
	\label{fig:grid_bol111}
\end{figure}

We fitted the spectra in XSPEC using an absorbed blackbody approach. Temperature and foreground \nh were both assumed to be the same during the two observations and only the respective normalisations were allowed to vary independently from each other. The relative stability of the spectral parameters is confirmed by a \xmm observation on 2008-07-18.26 UT (160 days later than 051138601), from which we extracted a spectrum of SS1 that can be fitted by a model with similar parameters (see below). The blackbody approach yields an acceptable $\chi^{2}_{r}$ = 1.39 for the best fit values of kT = $48^{+2}_{-3}$ eV and \nh = $2.3\pm0.1$ \hcm{21}. This fit is shown in Fig.\,\ref{fig:spec_bb_bol111} and the associated contour plot is given in Fig.\,\ref{fig:grid_bol111}. We computed the unabsorbed EPIC PN X-ray luminosities, in the range 0.2 - 1.0 keV, from the best fit model in XSPEC and used the best fit values to create in XSPEC fake spectra (command \texttt{fakeit}) to infer the energy conversion factors (ecf) for the \swift XRT (ecf$_{XRT}$) and the \chandra HRC-I (ecf$_{HRC-I}$). The ecf values are given in Table\,\ref{table:ss1_spec} and were used to convert our \swift XRT and \chandra HRC-I count rates to unabsorbed luminosities. All luminosities assume a distance to \m31 of 780 kpc \citep{1998AJ....115.1916H,1998ApJ...503L.131S} and are presented in Table\,\ref{table:ss1_xray}. Note, that for our first detection of SS1 in the \chandra HRC-I observation 8526 no photometry is possible due to the location of the source on the edge of the detector. The blackbody fit parameters and derived values, like luminosities for ObsID 0511380201, are given in Table\,\ref{table:ss1_spec}.

%

The unabsorbed X-ray luminosities inferred from the blackbody fit are in the order of \oergs{39}, and therefore significantly exceed the Eddington luminosity of the hydrogen rich atmosphere of a WD: $L_{\hbox{\rm Edd}} = 1.3 \times 10^{38} \left( \frac{M}{M_{\sun}}\right) \hbox{erg s$^{-1}$}$. The fact that blackbody fits to SSS spectra produce in general too high values of \nh and too low temperatures, and therefore too high luminosities, is well known \citep[see e.g.][and references therein]{1991A&A...246L..17G, 1997ARA&A..35...69K}, thus the values given in Table\,\ref{table:ss1_xray} define upper limits on the actual luminosity of SS1.

Blackbody fits are very simple approximations for SSS emission of novae and allow to compare parameters like effective temperature for different novae. However, as mentioned above, these fits are physically not realistic. Therefore, we tested fitting our low-resolution spectra with WD atmosphere models, that are based on more physical assumptions. We used a grid of synthetic ionizing spectra for hot compact stars from NLTE model atmospheres computed by \citet{2003A&A...403..709R}. These NLTE models are plane-parallel, in hydrostatic and radiative equilibrium and contain all elements from H to Ca \citep{1997A&A...320..237R}. The models were computed using the T\"ubingen Model-Atmosphere Package \citep[TMAP,][]{2003ASPC..288..103R}. Elemental abundances are fixed to either galactic halo ([X]\footnote{[]: log(abundance/solar abundance)} = [Y] = 0, [Z] = -1) or solar ([X] = [Y] = [Z] = 0) ratios. The grids of model atmosphere fluxes, as well as FITS tables which can be used in XSPEC, are available on-line\footnote{http://astro.uni-tuebingen.de/$\sim$rauch/}. The XSPEC tables contain temperatures and fluxes (binned to 0.1 \AA{} intervals) for fixed surface gravity ($\log g$) and elemental abundances. In our case, the available grid parameter space was restricted to models with $\log g = 9.0$, since only these tables include temperatures high enough to fit our spectra. This restriction clearly limits the significance of our best fits, since models with different surface gravity may have provided equally good fits with different parameters. Also note, that the assumptions of plane parallel and static are not physically realistic for a nova atmosphere. Furthermore, the spectral analysis is limited by the low energy resolution of the EPIC PN spectra, which are, due to the faintness of the source, the only available spectra with sufficient signal to noise ratio. Great caution should therefore be applied when interpreting the results of our test fits.

\begin{figure}
	\resizebox{\hsize}{!}{\includegraphics[angle=270,clip]{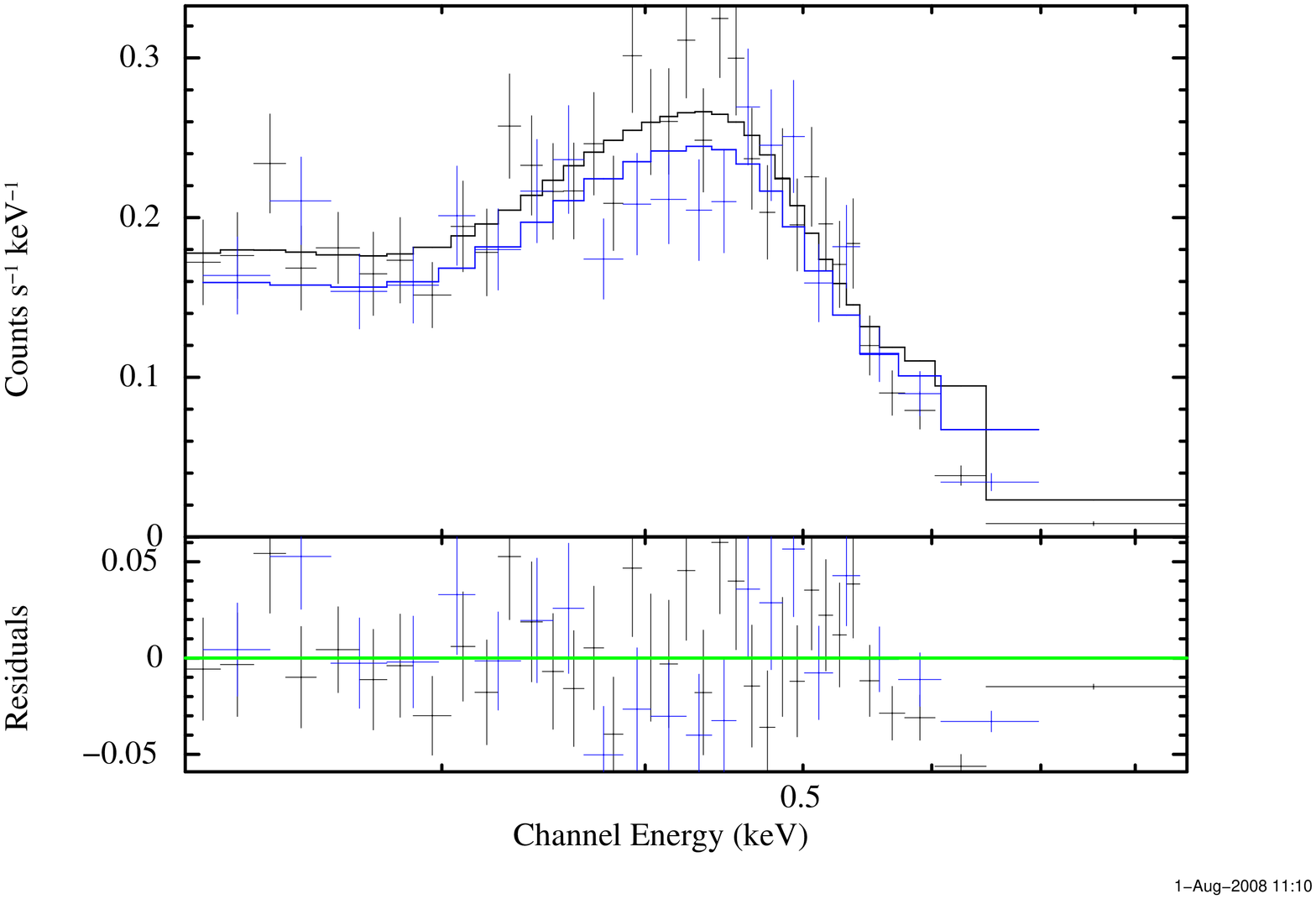}}
	\caption{\xmm EPIC PN spectra of SS1 from observations 0511380201 (\textbf{black}) and 0511380601 (\textbf{blue}) fitted with an WD atmosphere model with galactic halo abundances (\textbf{solid lines}).}
	\label{fig:spec_wd_bol111}
\end{figure}

We fitted the EPIC PN spectra with both halo and solar abundance ratios. Galactic halo abundances are more representative of the metallicity of the \m31 GCs \citep{2000ApJ...531L..29B}. Figure\,\ref{fig:spec_wd_bol111} gives the fit for halo abundances and shows that the atmosphere model fits the spectrum well only up to an energy of about 600 eV ($\chi^2$ dof$^{-1}$ = 1.01) but predicts too high flux at energies above, which is clearly indicated by the residuals. This is true for solar and halo abundances. However, this is no surprise since recent detailed X-ray spectroscopy of the SSS phase of galactic novae shows that these spectra are complex \citep[see e.g.][who used spectra obtained with the \xmm Reflection Grating Spectrometer (RGS) and the \chandra Low Energy Transmission Gratings Spectrometer (LETGS)]{2007ApJ...665.1334N}. Spectral models currently available may be not suitable for a correct fitting of SSS spectra of novae, especially with respect to the elemental abundances of a hot post-nova atmosphere.

In Table\,\ref{table:ss1_spec} we compare the parameters of the WD atmosphere model fits (fitted to the 0.2 - 0.6 keV range) to the parameters of the blackbody fit described above. We give unabsorbed X-ray luminosities($L_{x}$), that were computed for the range 0.2 - 1.0 keV in XSPEC, and bolometric luminosities ($L_{bol}$) as well as WD radii ($R$), inferred from the bolometric luminosities. The bolometric luminosity for the blackbody fit was directly computed from the normalisation of the spectra in XSPEC. For the WD atmosphere model we used an absorbed blackbody fit with the same temperature and foreground \nh to compute bolometric luminosity and WD radius. These values overestimate the actual luminosity and radius and are therefore upper limits. Energy conversion factors (ecf) for the \swift XRT (ecf$_{XRT}$) and the \chandra HRC-I (ecf$_{HRC-I}$) detectors were again computed in XSPEC, using fake spectra as for the blackbody fit, and are listed in Table\,\ref{table:ss1_spec} together with the ecf for the \xmm EPIC PN (ecf$_{PN}$).

The comparison of the fits shows that the WD atmosphere models give physically more plausible results. The high value of \nh which was required for the blackbody fit (three times the foreground absorption) is significantly reduced to a value which is just slightly above the foreground level of \nh = 0.7\hcm{21}. As a consequence, the super-Eddington luminosity derived from the blackbody fits is reduced to more realistic values for the atmosphere models. The connection of \nh and effective temperature can be seen in the confidence contours shown in Fig.\,\ref{fig:grid_bol111}. The effective temperature for the WD atmosphere is higher than the blackbody temperature and is revealing the H-burning hot layer. Still, one has to be cautious interpreting these models, because the elemental abundances are fixed and may be not realistic to describe abundances in a burning WD atmosphere. However, the consideration of model substructures, like absorption edges, already causes a reduction of the \nh necessary to adjust the model to the spectra. Truly physical models for supersoft emission from novae are strongly needed and should be validated using high resolution spectra obtained with current and future missions.

%
\begin{table}[ht]
\caption{Comparison of SS1 spectral best fit parameters and derived parameters for blackbody and WD atmosphere models with halo and solar abundances.}
\label{table:ss1_spec}
\begin{center}
\begin{tabular}{lrrr}
	\hline\noalign{\smallskip}
	\hline\noalign{\smallskip}
	Model & Blackbody & WD halo & WD solar\\
	(energy range [keV])  & ($0.2-0.8$) & ($0.2-0.6$) & ($0.2-0.6$)\\
	\noalign{\smallskip}\hline\noalign{\smallskip}
	kT (eV) & $48^{+2}_{-3}$ &  $61\pm1$ & $70\pm1$\\ \noalign{\smallskip}
	\nh ($10^{21}$ cm$^{-2}$) & $2.3\pm0.1$  &  $1.0\pm0.2$ & $1.0\pm0.2$\\ \noalign{\smallskip}
	$\chi^{2}_{r}$ & 1.39  & 1.01 & 1.06\\ \noalign{\smallskip}
	dof & 53  & 44  & 48\\ \noalign{\smallskip} \hline \noalign{\smallskip}
	$L_{x}$ ($10^{38}$ \hbox{erg s$^{-1}$}) & $10.6 \pm 0.2$ & $1.00 \pm 0.02$ & $0.88 \pm 0.02 $\\ \noalign{\smallskip}
	$L_{bol}$ ($10^{38}$ \hbox{erg s$^{-1}$}) & $28.7^{+0.2}_{-0.1}$ & $2.7 \pm 0.1$ & $1.5 \pm 0.1$\\ \noalign{\smallskip}
	$R$ (\power{9} cm) & $7.0^{+1.6}_{-0.7}$ & $1.21 \pm 0.07$ & $0.70 \pm 0.04$\\ \noalign{\smallskip}
	ecf$_{PN}$ (ct cm$^2$ erg$^{-1}$) & 5.0\tpower{9} & 5.5\tpower{10} & 6.1\tpower{10}\\ \noalign{\smallskip}
	ecf$_{HRC-I}$ (ct cm$^2$ erg$^{-1}$) & 2.6\tpower{9} & 3.2\tpower{10} & 3.7\tpower{10}\\ \noalign{\smallskip}
	ecf$_{XRT}$ (ct cm$^2$ erg$^{-1}$) & 7.4\tpower{8} & 8.7\tpower{9} & 9.9\tpower{9}\\ \noalign{\smallskip}
	 \hline
\end{tabular}
\end{center}
\noindent
Notes:\hspace{0.3cm} Luminosities and WD radii refer to the \xmm observation 0511380201. The unabsorbed X-ray luminosity $L_{x}$ is for the 0.2 - 1.0 keV range. Bolometric luminosities and WD radii for the WD atmosphere model are upper limits (see Sec.\,\ref{sect:sss_bol111} for details).
\end{table}
%

We checked recent \swift (ObsID 00037718001, starting at 2008-05-26.29 UT) and \xmm (ToO 0560180101, starting at 2008-07-18.26 UT) data and found that SS1 is still visible in both observations (see also Table\,\ref{table:ss1_xray} for details). Preliminary data analysis was published in \citet{2008ATel.1647....1P,2008ATel.1671....1P}. For \swift we used the blackbody fit described above and in Table\,\ref{table:ss1_spec} to convert the XRT count rate to unabsorbed luminosity. For \xmm we extracted an EPIC PN spectrum of SS1 as described above for the two earlier \xmm observations. The spectrum can be fitted with an blackbody fit with best fit values of kT = $43\pm8$ eV and \nh = $2.2^{+1.0}_{-0.7}$ \hcm{21} ($\chi^{2}_{r}$ = 1.52). These parameters are consistent with the blackbody fit described above and result in an unabsorbed X-ray luminosity (range of 0.2 - 1.0 keV) of ($8.9 \pm 0.5$) \ergs{38}. Both observations show that the X-ray luminosity of SS1 is slowly declining.

Supersoft spectra like the one of SS1 are typical for counterparts to optical novae and indicate that after the outburst there is hydrogen burning going on in the remaining envelope of the WD \citep[see][and references therein]{2005A&A...442..879P}. Therefore, we identify the supersoft X-ray transient in the globular cluster Bol 111 with the nova \nova reported by \citet{2007ApJ...671L.121S}. This optical nova was first detected on 2007 June 19.38 and was detected in X-rays the first time 141 days later on 2007 November 07.64. SS1 was not visible on 2007 July 22.55, which is 33 days after the first detection in the optical (see Table\,\ref{table:ss1_xray}).

\subsection{Supersoft source in Bol 194}
\label{sect:sss_bol194}
A second supersoft source (SS2) was detected in the same \chandra HRC-I observation as SS1 (ObsID 8526) as a new, bright X-ray transient \citep{2007ATel.1296....1H}. In the following observation (ObsID 8527) SS2 was again detected with similar brightness close to the edge of the HRC-I field of view (see Fig.\,\ref{fig:rotse_m31}). For the remaining three \chandra observations of the monitoring campaign the position of SS2 was outside the field of view due to the changing roll angle of the observations. Therefore, we used our \swift XRT ToO follow up observations (starting with ObsID 00031027001) to constrain the spectrum of SS2 and follow the X-ray light curve. The X-ray luminosities given in Table\,\ref{table:ss2_xray} show that this source faded much faster than SS1. Therefore, no \xmm observations of SS2 are available and we could use our \xmm data only to compute upper limits. Since SS2 also has a large off-axis position in the HRC-I field, we used the \swift observation 00031027001 to compute the position of the source. We found the following coordinates: RA(J2000) 00:43:45.3, Dec(J2000) +41:06:08.15. With a $1\sigma$ position error of $4\,\farcs6$ and a distance of $1\,\farcs1$ to the GC Bol 194 \citep[][00:42:45.20, +41:06:08.3]{2004A&A...416..917G} these coordinates are in good agreement with the position of Bol 194. See Fig.\,\ref{fig:xray_pos_err} for a comparison of optical (DSS POSS-II Red) and X-ray positions.

\begin{figure}
	\resizebox{\hsize}{!}{\includegraphics[angle=270,clip]{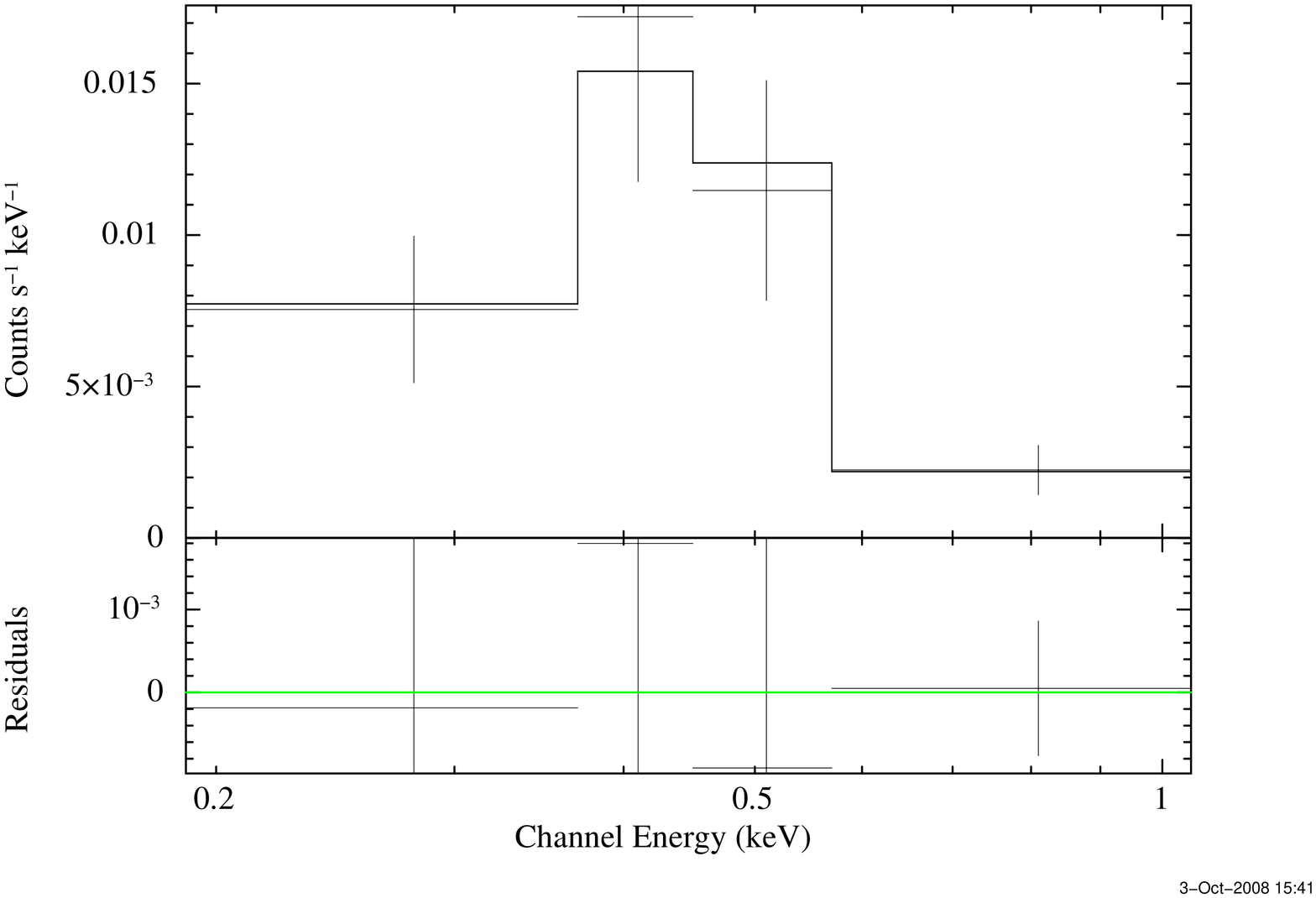}}
	\caption{\swift XRT spectrum of SS2 in Bol 194 from observation 00031027001 (7.3 ks exposure time) fitted with an absorbed blackbody fit (\textbf{solid line}).}
	\label{fig:spec_bol194}
\end{figure}

The \swift XRT spectrum of SS2 (ObsID 00031027001) only shows photons with energies below 750 eV (see Fig.\,\ref{fig:spec_bol194}). Therefore, we classified this source as a SSS. Since we have not many spectral counts for this source, we use Cash statistics for spectral modelling. The best spectral fit is an absorbed blackbody with best fit values of kT = $74^{+32}_{-23}$ eV and \nh = $\left(1.0^{+1.6}_{-0.9}\right)$ \hcm{21}. The best fit parameters differ from the preliminary data analysis published in \citet{2007ATel.1306....1H}, where we used a different abundance table in XSPEC \citep[angr;][]{1989GeCoA..53..197A} and $\chi^2$ statistics. Using the best fit values we computed the X-ray luminosity of SS2 to be (6.4 $\pm$ 1.0) \ergs{37}, the bolometric luminosity to be $10^{+7}_{-4}$ \ergs{37} and therefore the radius of the WD to be $5^{+11}_{-3}$ \tpower{8} cm. Unfortunately, the low count \swift XRT spectrum does not provide enough degrees of freedom to test the WD atmosphere model that we fitted to the \xmm EPIC PN spectra of SS1.

After our discovery, \citet{2007ATel.1328....1G} reported that they found SS2 to be present in a 5ks \chandra ACIS-S observation taken on 2007-11-03.18 UT, which is four days earlier than our first \chandra HRC-I detection. They confirm our classification of SS2 as a SSS and further report that the source is not visible anymore in a 5ks \chandra ACIS-S observation on 2007-11-27.16 UT ($\sim$ 24 days after the previous ACIS observation) with an 95 \% upper detection limit of $\sim$ 6 \cts{-4}. Note, that this observation was taken just three days after our Swift observation on 2007-11-24.34 UT. Additional \swift XRT observations by \citet{2007ATel.1334....1K} showed a re-brightening of the source on 2007-12-02.64. However, the source faded quickly again and was not found in follow-up \swift observations and our \xmm monitoring data. We re-analysed the \swift and \chandra HRC-I data, computed upper limits for the \xmm observations and present all available data on SS2 in Table\,\ref{table:ss2_xray}. We took the count rates for detections and upper limits from \citet{2007ATel.1334....1K} to compute luminosities in XSPEC using our spectral model. For the \swift observations 00031027004/5 we give 95\% upper limits on the SS2 luminosity. These values are comparable to actual luminosities obtained for earlier \swift observations, an effect which is due to longer effective source exposure times for the earlier observations. X-ray luminosities for the three instruments were computed using XSPEC and the blackbody fit inferred from the \swift spectrum.

Analogous to SS1 and its possible connection to \nova in Bol 111, the supersoft spectrum and the transient light curve could indicate that SS2 is the X-ray counterpart of a recent optical nova in the GC Bol 194. The following section describes the optical data analysis and the constraints that we can put on a possible counterpart nova in Bol 194.

%
\section{Search for an optical nova counterpart of the supersoft source in Bol 194}
\label{sec:obs_optical}
%
As there was no optical nova reported in Bol 194 we searched the recent optical data available to us for indications of a nova outburst in the GC. Note, that the constant background light from the GC makes it difficult to detect nova outbursts. However, difference imaging is a successful method in this context and was used to analyse a big part of our optical data set. In the following we describe the observational setup and the analysis of the optical data.

\subsection{Optical observations}
The majority of our optical data was obtained in the context of the
Texas Supernova Search (TSS; \citealt{2006quimby_phd}). The TSS employed
the 0.45-m ROTSE-IIIb telescope at the McDonald observatory in Texas,
and the data presented here are supplemented by its twin, the
ROTSE-IIId telescope, which is located at the Turkish National
Observatory at Bakirlitepe, Turkey. The ROTSE-III system is described
in \citet{akerlof03}. The $1\fdg85 \times 1\fdg85$ field of view
covered by the unfiltered, 2k$\times$2k Marconi CCD encompasses
practically all of \m31's light in a single exposure (see
Fig. \ref{fig:rotse_m31}), and additional
overlapping fields to the northeast and southwest add to the haul. Beginning in November 2004, these 3 fields were
imaged several times nightly as weather and season allowed.

Our second monitoring program for optical novae uses the robotic 60 cm telescope with an E2V CCD (2kx2k) Livermore Optical Transient Imaging System \citep[Super-LOTIS,][]{2008AIPC.1000..535W} located at Steward Observatory, Kitt Peak, Arizona, USA. Starting in October 2007 the telescope was used in every good night to monitor the bulge of M\,31. Using four Super-LOTIS fields (field of view: $17\arcmin \times 17\arcmin$) these observations cover an area of $\sim 34\arcmin \times 34\arcmin$ centered on the core of M\,31 with a pixel scale of 0.496\arcsec/pixel and a typical limiting magnitude of 19.0 mag, using a Johnson R filter. The reduction of the data is done by a semi-automatic routine and the astrometric and photometric calibration uses the \m31 part of the Local Group Survey \citep[LGS,][]{2006AJ....131.2478M}. Typical values of Super-LOTIS astrometric and photometric $1\sigma$ accuracies are $0\,\farcs25$ and 0.25 mag, respectively, averaged over the whole magnitude range. The southeast Super-LOTIS field covers Bol 194 and the astrometric and photometric $1\sigma$ accuracies for this particular object ($\sim 16.8$ mag) are $0\,\farcs11$ and 0.06 mag, respectively.

Additionally, in this research our optical data set is supplemented by archival data from K. Hornoch obtained at telescopes in  Lelekovice (Newtonian focus of 350/1658 mm telescope, CCD camera G2CCD-1600,
Kron-Cousins R filter, 1.12\arcsec/pixel, FOV 28.7\arcmin x19.1\arcmin) and Ond\v{r}ejov (primary focus of 650/2342 mm telescope, CCD camera G2CCD-3200, Kron-Cousins R filter, 1.20\arcsec/pixel, FOV 21.8\arcmin x14.7\arcmin).
Standard reduction procedures for raw CCD images were applied (dark and bias
subtraction and flat-field correction) using SIMS\footnote{\tt http://ccd.mii.cz/}
and  Munipack\footnote{\tt http://munipack.astronomy.cz/}
programs. Reduced images of the same series were co-added to improve
the S/N ratio (total exposure time varied from 600s up to 1800s).
The gradient of the galaxy background of co-added images was flattened
by the spatial median filter using SIMS. These processed images were used
for aperture photometry, carried out in GAIA\footnote{\tt
http://www.starlink.rl.ac.uk/gaia}. 
Relative photometry was done using brighter field stars which were calibrated using standard Landolt fields.
The 1$\sigma$ measurement uncertainties were low ($\sim 0.03$ mag), thanks to the long exposure times, the brightness of Bol 194, and its location far from the high surface brightness levels found near the center of \m31.

\subsection{Optical data analysis}
We searched the optical data of Bol 194 obtained by the monitoring programs described above for a significant optical excess that could indicate a nova outburst in the GC. The significance for the Super-LOTIS data and for Hornoch's data is defined by applying a $3\sigma$ clipping to the light curves of Bol 194 and judging every two neighboured data points that lie outside the final $\pm3\sigma$ range as possible indications of a nova outburst.

For the ROTSE-III data, we used the PSF-matched image subtraction code
developed by the Supernova Cosmology Project~\citep{perlmutter99} to
search for residual light from an optical nova. With this method the outburst of nova \nova in Bol 111 was detected \citep{2007ApJ...671L.121S}. We first constructed a
deep reference image by co-adding 100 ROTSE-IIIb images obtained
between December 2004 and June 2006. We then convolved this image to
match the PSF of each ROTSE-III image, subtracted off this template
light, and searched for any point sources coincident with Bol 194. We
attempted to measure any residuals at this location with the DAOPHOT
PSF-fitting routines (\citealt{stetson87}; ported to IDL by
\citealt{landsman89}).

\begin{figure}
	\resizebox{\hsize}{!}{\includegraphics[angle=270]{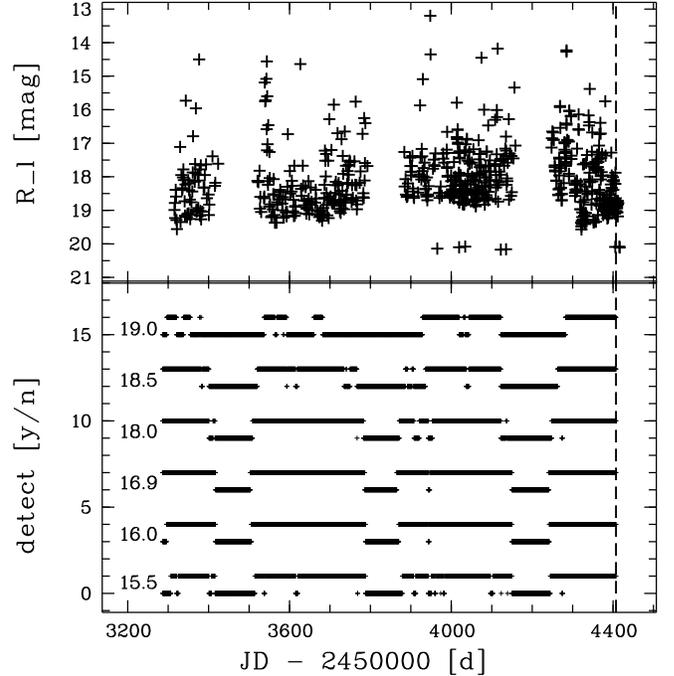}}
	\caption{\textbf{Top}: Limiting magnitudes of all optical data since 2004 November. \textbf{Bottom}: simulated detection (upper points) or non detection (lower points) for novae with indicated peak magnitudes. In both panels the vertical dashed line shows the date of the first detection of SS2 in X-rays.}
	\label{fig:bol194_all}
\end{figure}

These procedures did not lead to the discovery of an optical nova counterpart of SS2 in Bol 194. However, our optical monitoring data of the M\,31 bulge region allows us to put strong constraints on the outburst date of any nova that may have occurred in Bol 194. Figure \ref{fig:bol194_all} includes all our optical data of Bol 194 from November 2004 until 2007-11-08, which is 5 days after the first detection of SS2 in X-rays on 2007-11-03. We show the minimum detectable magnitude of a possible optical nova occurring in Bol 194 during this period.

To calculate the limiting magnitudes of the ROTSE-III observations, we
measured the noise on the subtracted frames in annuli centered on the
location of Bol 194. The limits reported here correspond to the flux
required of a point source to be detected at the $4\sigma$ level. The
magnitude scale was calibrated against the USNO-B1.0 R2 measurements \citep{2003AJ....125..984M}.

In contrast to this procedure, the limiting magnitude for the Super-LOTIS data and for Hornoch's data is computed as follows. At the distance of \m31 the light of a nova in a GC would blend with the light of the GC itself. Therefore, we have to take into account the intrinsic magnitude of the cluster ($\sim 16.8$ mag in the Super-LOTIS data) and compute the resulting magnitude using
\begin{displaymath}
R_t = -2.5\,log_{10}(10^{-0.4\, R_{gc}} + 10^{-0.4\, R_n}),
\label{eqn:summag}
\end{displaymath}
with $R_t$, $R_{gc}$, and $R_{n}$ being the total magnitude, the magnitude of the GC, and the magnitude of the nova, respectively. Thus, for the Super-LOTIS data and for Hornoch's data, the minimum detectable magnitude $R_{l}$ of a nova is defined as the magnitude that leads to a significantly brighter $R_t$. The significance criterion used here is:
\begin{displaymath}
R_{t} < R_{gc} - 3\sigma_{R_{gc}},
\label{eqn:signif}
\end{displaymath}
with $3\sigma_{R_{gc}}$ being the 3$\sigma$ photometric standard error for Bol 194 ($\sim 0.16$ mag for the Super-LOTIS data reduction and $\sim 0.09$ mag for Hornoch's data). These measurement errors are mean values derived from our light curves of Bol 194. Since we used a $3\sigma$ clipping method to search for a nova outburst in Bol 194, we assume that the mean photometric errors given for both instruments would correspond to the final 3$\sigma$ range in the case of a nova outburst-modified light curve. Therefore, we compute $R_{l}$ as follows:
\begin{displaymath}
R_{l} = -2.5\, log_{10}(10^{-0.4\,(R_{gc} - 3\sigma_{R_{gc}})} - 10^{-0.4\, R_{gc}}).
\label{eqn:limmag}
\end{displaymath}

In order to judge the time coverage of our data set, we simulated the outburst of novae with different peak magnitudes and decay times in Bol 194 on any day between 2004 October and 2007 November and checked whether we would have detected the outburst according to the limiting magnitudes, or not. We simulated novae with peak magnitudes of 15.5, 16.0, 16.9, 18.0, 18.5 and 19.0 mag in the R band and decay times of 6, 15, 29, 63, 102 and 164 days, respectively. The peak magnitude of nova \nova in Bol 111 as given in \citet{2007ApJ...671L.121S} is 16.9 mag. The decay time $t_2$ is defined as the time in days the nova luminosity needs to drop 2 mag below peak luminosity \citep{1964gano.book.....P}. To compute this value, we used the maximum magnitude versus rate of decline (MMRD) relationship given by \citet{1995ApJ...452..704D}, for $t_2 \leq 50$ d:
\begin{displaymath}
M_{V,max} = -7.92 - 0.81 \arctan \left( \frac{1.32 - \log t_2}{0.23}\right),
\label{eqn:mmrd1}
\end{displaymath}
and by \citet{1988ASPC....4..114C}, for $t_2 > 50$ d:
\begin{displaymath}
M_{V,max} = 2.41 \log t_2 - 10.70.
\label{eqn:mmrd2}
\end{displaymath}

We used this combination of the two different MMRDs according to \citet{1997ApJ...487..226S} in order to compensate for the breakdown of the MMRD from \citet{1995ApJ...452..704D} at large values of $t_2$. Both MMRDs give the relation of the decay time of a nova to its maximum magnitude in the $V$ band. The color relation for novae at maximum light is $(B-V)_0 = 0.23\pm0.06$ \citep{1987A&AS...70..125V}. This fact implies for novae at maximum light a spectral type close to F0 and therefore a $(V-R)_0 \sim 0.30$ \citep[see e.g.][]{1976asqu.book.....A} which we used to match both MMRDs to our $R$ band data.

For our simulations we used a \m31 extinction free distance modulus of $\mu_0 = 24.38$ mag \citep{2001ApJ...553...47F}. We applied a reddening towards \m31 of $E(B-V) = 0.062$ \citep{1998ApJ...500..525S} and a foreground extinction in the R band of $A^{f}_{R} = 0.17$ mag, obtained via the NASA Extragalactic Database (NED). We estimated the internal absorption of \m31 to be $A^{i}_{R} \sim 0.11$ mag, using $A^{i}_{pg} \sim 0.20$ mag \citep{1989AJ.....97.1622C} and assuming that $A_{pg} \sim A_{B}$.

The results of these simulations are shown in the top panel of Fig.\,\ref{fig:bol194_all}, with the upper parts of the curves indicating the periods over which each simulated nova would have been detected. The figure visualises the fact that our monitoring is efficient in the sense that the only periods in which we would not have been able to detect the most novae, are the periods when \m31 is not observable due to its annual visibility. These periods, computed for a nova with parameters of \novak, and the estimated constraints on the delay time between optical outburst and detection in X-rays are shown in Table\,\ref{table:nova_bol194}. Note also, that a bright nova (peak luminosity = 15.5 mag) would have been missed more likely than a fainter nova. This is due to the faster decline of luminosity for the brighter novae and good detection limits of our observations down to $\sim$ 19 mag.

%

%
\begin{table}[ht]
\caption{Optical constraints on a nova in Bol 194}
\label{table:nova_bol194}
\begin{center}
\begin{tabular}{lccll}\hline\hline
	Date1 [UT]$^a $ & Date2 [UT]$^a $ & RJD1 [d]$^b $ & RJD2 [d]$^b $ & Delay [d]$^c $ \\ \hline
	2005-02-15 & 2005-05-12 & 3417 & 3503 &  990 - 904\\
	2006-02-19 & 2006-05-09 & 3786 & 3865 & 621 - 542\\
	2007-02-18 & 2007-05-20 & 4150 & 4241 & 257 - 166\\ \hline
\end{tabular}
\end{center}
\noindent
Notes:\hspace{0.3cm} $^a $: Start date (date1) and end date (date2) of the non-\\
\hspace*{1.4cm} detection periods\\
\hspace*{1.1cm} $^b $: The same as in $^a $ but in RJD = JD - 2\,450\,000\\
\hspace*{1.1cm} $^c $: Periods of possible delay time between nova outburst\\
\hspace*{1.4cm} and first detection in X-rays\\
\end{table}
\noindent
%

\section{Discussion}
\label{sec:discuss}
%

%
\subsection{Summary of observational data}
We present in this paper the first two SSSs found in \m31 GCs. The source SS1 in Bol 111 can be identified with \novak, the very first nova found in a \m31 GC in June 2007 \citep{2007ApJ...671L.121S}. After 137 days (November 2007) we detected supersoft X-ray emission from this source and it was still active in our last \xmm observation in July 2008. The supersoft spectrum and the transient nature of SS2 in Bol 194 could also indicate a recent nova outburst. However, we could not find evidence for an nova counterpart in optical monitoring data of this GC. We give constraints on a possible nova outburst in Bol 194, based on the detection limits of our optical observations and simulated nova outbursts. The simulations show that most novae would have been detected in Bol 194, except in the times when \m31 could not be observed due to it's annual visibility. We computed the time spans in which a nova like \nova in Bol 194 could have had an unobserved outburst. The associated delay times, given in Table\,\ref{table:nova_bol194}, between optical outburst and X-ray switch-on are physically feasible, since similar delays have been observed in other supersoft nova counterparts \citep[see][]{2007A&A...465..375P}. Therefore, a nova outburst in Bol 194, that is connected to SS2, cannot be ruled out.

\subsection{Globular cluster nova rate - the optical point of view}
\label{sec:rate_opt}
The statistics of novae in GCs of other galaxies is still poor. \citet{1992BAAS...24.1237T} conducted a search in H$\alpha$ of over 200 \m31 GCs for nova eruptions and found nothing over an effective survey time of one year. This results in an upper limit for the \m31 GC nova rate of 0.005 novae yr$^{-1}$ GC$^{-1}$. Another search for novae in 54 \m31 GCs was done by \citet{1990PASP..102.1113C} based on the \m31 H$\alpha$ survey data of \citet{1987ApJ...318..520C,1990ApJ...356..472C}. Over a mean effective survey time of approximately two years no indications for a nova outburst in one of the GCs was found. From one nova found in a GC of the giant elliptical galaxy M\,87 \citet{2004ApJ...605L.117S} derive a rate of 0.004 novae yr$^{-1}$ GC$^{-1}$ for the whole system of 1057 known M\,87 GCs (4.2 novae yr$^{-1}$).

There are about 500 GCs known that belong to \m31 (482 confirmed GCs, according to the Revised Bologna Catalogue\footnote{\texttt{http://www.bo.astro.it/M31/}} \citep[V.3.5, March 2008;][]{2004A&A...416..917G,2005A&A...436..535G,2006A&A...456..985G,2007A&A...471..127G}). This corresponds approximately to a stellar mass of about 5 \tpower{9} M$_{\sun}$. Given a rate of 2.2 novae yr$^{-1}$ (\power{10} M$_{\sun}$)$^{-1}$, which is observed in old systems like elliptical galaxies (data derived from Table\,3 of \citet{1994A&A...287..403D} and Equation (1) of \citet{2005A&A...433..807M}), we derive a rate of about 1.1 novae yr$^{-1}$ for the \m31 GC system, which corresponds to about 0.002 novae yr$^{-1}$ GC$^{-1}$. This figure is comparable (within a factor of 2) with the observed rate (0.004) of the M\,87 GC system obtained by \citet{2004ApJ...605L.117S} and suggests that the discovery of only one nova in the \m31 GC system \citep{2007ApJ...671L.121S} over about a century of observations may be the result of a strong observational bias \citep[see][]{2007ApJ...671L.121S}.

\subsection{Globular cluster nova rate - the X-ray point of view}
\label{sec:rate_xray}
While the detection of CNe in GC in the optical is strongly hampered by the light from the GC itself, the detection of supersoft emission from a hydrogen-burning post-nova atmosphere is not affected by this. However, the presence of other X-ray sources in a GC, such as LMXBs, may create a similar problem for the detection of a SSS in that GC. In the following, we discuss the impact of our results on the rate of novae for the \m31 GCs.

If we assume that both SS1 and SS2 are post-novae and assume the duration of the SSS phase to be of the order of 1 yr then we find a nova rate of 0.015 novae yr$^{-1}$ GC$^{-1}$ for the $\sim$ 130 GCs from \citet{2007A&A...471..127G} in the field of view of our \chandra HRC-I observations (see also Fig.\,\ref{fig:rotse_m31} for the \chandra field). The duration of the SSS phase of novae is a critical parameter for a correct estimate of the nova rate. Although there are some novae known to have much longer SSS phases than 1 yr \citep[see e.g.][]{1993Natur.361..331O} it is likely \citep[see][]{2007A&A...465..375P} that novae with short SSS phases were just not found in the past due to selection effects. Also note, that the SSS phase of SS2 lasted only $\lesssim$ 4 months and therefore the assumed duration of 1 yr may still be a quite conservative estimate.

The nova rate computed using the X-ray data on the SSSs is about a factor 10 larger (after assuming the duration of the SSS phase to be 1 yr) than the nova rate estimated by assuming that GCs produce novae with a rate comparable to the old stellar populations in Ellipticals. The match between the rates in X-ray and in optical is obtained for a duration of the SSS phase of about a decade. Such a long time has been observed indeed \citep{1993Natur.361..331O}, but it does not seem to be the rule \citep{2007A&A...465..375P}.

The results based on the small size of our sample of GC SSSs are clearly dominated by Poissonian statistics. In this case, the discovery of two GC novae would still be consistent with a nova rate of $\sim$ 0.002 novae yr$^{-1}$ GC$^{-1}$ (95\% confidence level) and $\sim$ 0.004 novae yr$^{-1}$ GC$^{-1}$ (99.87\% confidence level), respectively. Both values are below the upper limit of 0.005 novae yr$^{-1}$ GC$^{-1}$ estimated by \citet{1992BAAS...24.1237T} from optical observations.

Furthermore, we want to emphasize the fact that not all optical novae in the Galaxy and in \m31 were detected as SSSs. Novae with supersoft X-ray emission are a subset, with yet unknown size, of the whole nova population. The nova rates deduced from X-ray observations are therefore lower limits on the actual nova rates.

\subsection{Conclusions}
The nova rate in the \m31 GC system may be larger by about one order of magnitude than expected in old stellar systems like giant ellipticals. This fact implies the existence of some other (dynamical) mechanism which acts inside the GCs and is capable to raise the fraction of CN systems naturally produced by stellar evolution. It is well known that LMXBs are highly overabundant in GCs, with respect to the rest of a galaxy \citep[see e.g.][and references therein]{2005PASP..117.1236F}. This is explained by dynamical effects like tidal captures of low-mass main-sequence stars by neutron stars \citep[see also][]{1975ApJ...199L.143C,1975MNRAS.172P..15F}. One would expect similar effects for WDs \citep{1983Natur.301..587H}, but although large numbers of cataclysmic variables (CVs) have been found in Galactic GCs, the number of novae detected in GCs is still small \citep[see][and references therein]{2004ApJ...605L.117S}.

Further, we note that novae with short SSS states seem to be an important contributor to the SSS population of galaxies \citep[see also][]{2007A&A...465..375P}. \xmm observations of \m31 in the past \citep{2008A&A...480..599S,2005A&A...434..483P,2004ApJ...616..821T} did not detect SSS with positions that agree with GC positions. This might in part be due to the fact that the individual observations of these monitorings were separated by half a year or more. The discoveries of SS1 and SS2 are the first results of a new monitoring strategy, that uses \xmm and \chandra observations that are separated just by 10 days. Future observations that follow this strategy may increase the statistics for SSS in \m31 GCs. If future surveys of \m31 will be able to find between six to eight new GC novae per year, the existence of a nova rate excess, of the size reported above, will be proven at 95\% and 99.87\% confidence level, respectively.

Finally, given the possible link between SSS and type Ia supernovae (SNe-Ia) \citep[e.g.][]{1994ApJ...423L..31D,1994ApJ...423..274D} if a high rate for SSS in GCs will be confirmed by future surveys, it may not be inconceivable to expect detection of SNe-Ia in GC systems around giant ellipticals or bulge dominated galaxies.

\begin{acknowledgements}
We wish to thank the referee, Jan-Uwe Ness, for his constructive comments which helped to improve the manuscript significantly. The X-ray work is based in part on observations with \xmmk, an ESA Science Mission with instruments and contributions directly funded by ESA Member States and NASA. The \xmm project is supported by the Bundesministerium f\"{u}r Wirtschaft und Technologie / Deutsches Zentrum f\"{u}r Luft- und Raumfahrt (BMWI/DLR FKZ 50 OX 0001), the Max-Planck Society and the Heidenhain-Stiftung. We would like to thank the \swift team for the scheduling of the ToO observations. This research has made use of the NASA/IPAC Extragalactic Database (NED) which is operated by the Jet Propulsion Laboratory, California Institute of Technology, under contract with the National Aeronautics and Space Administration. This study made use of the Digitized Sky Surveys which were produced at the Space Telescope Science Institute under U.S. Government grant NAG W-2166. M. Henze and H.S. acknowledge support from the BMWI/DLR, FKZ 50 OR 0405. G.S. acknowledges support from the BMWI/DLR, FKZ 50 OR 0405, and from grants AYA2008-04211-C02-01 and AYA2007-66256. M. Hernanz acknowledges support from grants ESP2007-61593 and 2005-SGR00378.
\end{acknowledgements}

\bibliographystyle{aa}

\end{document}